\documentclass[12pt]{JHEP3}
\usepackage{amssymb}
\usepackage{amsfonts}
\usepackage{epsf}

\font\mybb=msbm10 at 11pt
\def\bb#1{\hbox{\mybb#1}}
\def\bZ {\bb{Z}}

\renewcommand{\a}{\alpha}
\renewcommand{\b}{\beta}

\renewcommand{\d}{\delta}
\newcommand{\rmd}{{\rm d}}
\newcommand{\m}{\mu}
\newcommand{\n}{\nu}

\def\be{\begin{equation}}
\def\ee{\end{equation}}
\def\bea{\begin{eqnarray}}
\def\eea{\end{eqnarray}}
\def\ba{\begin{array}}
\def\ea{\end{array}}
\def\bi{\begin{itemize}}
\def\ei{\end{itemize}}

\title{{\bf $SL(2,Z)$ Action on
Three-Dimensional CFTs and Holography}}

\author{
Robert G. Leigh\\
        Department of Physics\\
        University of Illinois at Urbana-Champaign\\
        Urbana, IL 61801 USA\\
  Email: \email{rgleigh@uiuc.edu}}
\author{Anastasios C. Petkou\\
CERN Theory Division, \\
CH-1211 Geneva 23,
Switzerland\\
Email: \email{tassos.petkou@cern.ch}
}

\keywords{AdS-CFT Correspondence, Duality in Gauge Field Theories}
\preprint{CERN-TH/2003-215\\ ILL-(TH)-03-08}

\abstract{ We show that there is a natural action of $SL(2,\bZ)$ on the
two-point functions of the energy momentum tensor and of higher-spin
conserved currents in three-dimensional CFTs. The dynamics behind the
$S$-operation of $SL(2,\bZ)$ is that of an irrelevant current-current
deformation and we point out its similarity to the dynamics of a wide
class of three-dimensional CFTs. The holographic interpretation of our
results raises the possibility that many three-dimensional CFTs have
duals on AdS$_4$ with $SL(2,\bZ)$ duality properties at the linearized level. 
}

\begin{document}

\section{Introduction}

Conformal field theories in spacetime dimensions $d>2$ are relatively
rare entities. In the particularly interesting case of four spacetime
dimensions, it seems possible that non-trivial CFTs require
supersymmetry, while the majority of them do not have a simple
Lagrangian formulation. Nevertheless, it has been known for a long time
that in three dimensions there exists a number of non-trivial CFTs that
do not necessarily require supersymmetry, while at the same time they do
have a simple Lagrangian formulation that allows the evaluation of
dynamical quantities like anomalous dimensions. Such three-dimensional
CFTs have a wealth of interesting perturbative as well as
non-perturbative properties that appear to be physically relevant as
they describe universality classes of real statistical systems (for a
recent review containing a large list of references see \cite{MZJ}).

The recent success of ideas related to holography leads naturally to
question whether the rich web of three-dimensional CFTs could be related
to quantum theories on (A)dS$_4$ spaces. This simple question turns out
to be unexpectedly important in view of the recent interest in quantum
field theory on four-dimensional spaces with cosmological constant.
There is, however, an additional reason to be interested in the
holographic properties of three-dimensional CFTs. It is known that in
many of the latter theories one has control on both the weak-coupling
(free field theory limit) and the strong-coupling regimes. This property
provides the right framework to study the holography of free field
theories, a subject that has recently attracted attention
\cite{Gopakumar, Bianchi, Sundborg}.

A bold proposal for a holographic dual of a three-dimensional CFT was
first made in \cite{KP1} for the Critical $O(N)$ Vector
model\footnote{Earlier ideas for a connection between free CFTs and 
Higher-Spins are recorded in \cite{SS0}.} and and its
connection with a Higher-Spin theory\footnote{For a recent work on
Higher-Spins with an
extensive reference list see \cite{Vasiliev,Sagnotti}.} on AdS$_4$. Soon
thereafter the Higgs mechanism giving masses to the Higher-Spins in
AdS$_4$ was discussed \cite{GPZ}.  In
\cite{Tassos1} we have argued that such a proposal could 
lead to the reconstruction of a consistent quantum theory on AdS$_4$.
Moreover, it was further realized in \cite{LP,SS} that both the Critical
$O(N)$ Vector model and the fermionic $O(N)$ Gross-Neveu model can be
contained in the holographic dual of a unique ${\cal N}=1$
supersymmetric Higher-Spin theory on AdS$_4$.

An interesting yet apparently unrelated observation was
recently made by Witten in \cite{Witten} (see also
\cite{LR,Lutken,SK}). There it was shown that there 
is a natural $SL(2,\bZ)$ action on  three-dimensional CFTs with $U(1)$
conserved currents. This action consists of a sequence of operations,
described in detail in \cite{Witten}, that lead from a given CFT to its 
dual. The relevant point for us is that the $S$- and $T$-operations of
$SL(2,\bZ)$ act on the two-point function of the conserved spin-1
current of the initial CFT and produce the two-point function of the 
spin-1 conserved current of
the dual CFT. Quite intriguingly, it was further argued in \cite{Witten}
that the $SL(2,\bZ)$ action on the three-dimensional CFT may be viewed
as  the holographic image of the well-known $SL(2,\bZ)$ duality of pure
electrodynamics on AdS$_4$.

In this work we show that the $SL(2,\bZ)$ action naturally extends to
two-point functions of the energy momentum tensor and of higher-spin
conserved currents in three-dimensional CFTs. In particular, we show
that the $S$-operation of $SL(2,\bZ)$ on two-point functions is
implemented by coupling the corresponding current to an external field,
evaluating its induced propagator and reading from the latter the
two-point function of an appropriately defined dual conserved current.
We present the calculation for the energy momentum tensor (spin-2
current) in some detail and sketch the corresponding calculation for
higher-spin conserved currents. The $T$-operation of $SL(2,\bZ)$ acts as
in \cite{Witten} by shifting the coefficient of the conformally
invariant contact term which appears in the two-point functions of all
of these conserved currents.

Moreover, we show that the $S$-operation on the two-point functions of
conserved currents may be implemented dynamically by certain
double-trace deformations.\footnote{In a slight abuse of terminology, we
keep here the notion of ``double-trace'' composite  operators in
theories with elementary $O(N)$ vector fields in par with recent
literature.} We point out the similarity of these double-trace
deformations to the dynamics of some well-known three-dimensional CFTs,
such as the $O(N)$ Vector and Gross-Neveu models and the Thirring model.

Then we look for a holographic interpretation of our $SL(2,\bZ)$ action.
This leads us naturally to consider theories on AdS$_4$ that are
generalizations of the self-dual pure electrodynamics, at least at the
linearized level. For gravity, we argue that such a theory may be given
by the old MacDowell-Mansouri action \cite{MM} with the addition of the
topological Pontryagin term that plays the role of a $\theta$-term
\cite{Mexicans,Hull}. We show that this action gives rise to the two
coefficients in the two-point function of the boundary energy momentum
tensor, in the same way as pure electromagnetism gives rise to the two
coefficients in the two-point function of the boundary spin-1 current.
For higher-spins, the relevant AdS$_4$ action may be the recently
discussed action of Vasiliev \cite{Vasiliev2} with the addition of the
appropriate Pontryagin term for higher-spin curvatures of $SO(3,2)$.
Therefore, we raise the possibility that a number of three-dimensional
CFTs may have AdS$_4$ duals with $SL(2,\bZ)$ duality properties, at least at
the linearized level. This also emphasizes the intimate relation of
$SL(2,\bZ)$ duality to the holography of free field theories.

The paper is organized as follows. In section 2 we demonstrate the
$S$-operation on two-point functions of higher-spin conserved currents.
In section 3 we show that the $S$-operation can be implemented by
double-trace deformations, in close analogy to the dynamics
of some well-known three-dimensional CFTs. In section 4 we discuss the
holographic interpretation of our results and in particular a
gravitational action that may induce our $SL(2,\bZ)$ action in the
boundary. We conclude and discuss further implications of our results in
section 5.

\section{$SL(2, Z)$ action on two-point functions of conserved currents}

\subsection{Review of the $U(1)$ current case}

We begin with a review of Witten's result\cite{Witten} (see also
\cite{BD}) for the two-point function of a conserved spin-1
current.  With the definition
\be
\label{Pimn}
\Pi^{\m\n}(p) =p^\m p^\n -\d^{\m\n}p^2,
\ee
the momentum space two-point  function of a spin-1 current in a three-dimensional
CFT  has the general form
\be
\label{2ptJW}
\langle J^\m(p)J^\n(-p)\rangle\equiv J^{\m\n}(p)
=C_J\frac{\Pi^{\m\n}(p)}{\sqrt{p^2}}
+W_J\epsilon^{\m\n\rho}p_\rho\,.
\ee
In coordinate space this takes the form
\be
\label{2ptJx}
\langle J^\mu(x) J^\nu (0)\rangle=C_J\Pi^{\mu\nu}(i\partial)
\left(\frac{1}{2\pi^2x^2}\right)+W_J\epsilon^{\mu\nu\rho}i\partial_\rho \delta^{(3)}(x) \,.
\ee
The second term in (\ref{2ptJx}) is a conformally invariant contact term
that is special to three dimensions. The effect of the $S$-operation
\cite{Witten} is to transform the two parameters $C_J$ and $W_J$  as
\be
\label{JStransf}
C_J\rightarrow \frac{C_J}{C_J^2 +W_J^2}\,,\,\,\,\,\, W_J\rightarrow
-\frac{W_J}{C_J^2 +W_J^2}\,.
\ee
The $T$-operation is a rather trivial shift in the value of $W_J$ by an
integer. The important result of \cite{Witten} is that these $S$- and
$T$-operations generate the modular group $SL(2,\bZ)$. In what follows
we concentrate on the non-trivial $S$-operation.

A  simple way to obtain the result (\ref{JStransf}) is to couple
$J^\m$ to an external vector field $A_\m$ and integrate
elementary fields (assuming the existence of an appropriate Lagrangian
description), to obtain the effective action
\begin{equation}
\label{effectJ}
\frac12\int \frac{d^3p}{(2\pi)^3} \ A_\m(p)J^{\m\n}(p)A_\n(-p)+\ldots
\end{equation}
This would yield an effective propagator for $A_\m$; however
$J^{\m\n}(p)$ is not directly invertible so if we want to proceed we
should gauge fix, adding
\begin{equation}
\frac12\int \frac{d^3p}{(2\pi)^3}\ A_\m(p)\left[ C_J(\xi-1)
\frac{p^{\m}p^{\n}}{\sqrt{p^2}}\right]A_\n(-p)\,.
\end{equation}
This leads to the addition of the quantity in square brackets to
$J^{\m\n}(p)$ as
\begin{equation}
J_{\xi}^{\m\n}(p)=C_J\frac{1}{\sqrt{p^2}} 
\left(\xi p^\m p^\n-p^2\eta^{\mu\nu}\right)+W_J\epsilon^{\m\n\rho}p_\rho \,.
\end{equation}
The tensor above has the inverse
\begin{equation}
M^{\xi}_{\m\n}(p)=\frac{1}{p^2}\left[ \frac{C_J}{C_J^2+W_J^2}
\frac{\Pi_{\m\n}(p)}{\sqrt{p^2}}-\frac{W_J}{C_J^2+W_J^2}\epsilon_{\m\n\rho}p^\rho+
\frac{1}{C_J(\xi-1)}\frac{p_\m p_\n}{\sqrt{p^2}}\right]\,,
\end{equation}
which is the two-point function of $A_\m$. 
Defining then the dual conserved current as
$\hat{J}^\m(p)=i\epsilon^{\m\n\rho}p_\n A_\rho(p)$ we
find its two-point function to be
\begin{equation}
\label{2ptdualJ}
\langle\hat{J}^\m(p)\hat{J}^\n(-p)\rangle=
\frac{C_J}{C_J^2+W_J^2}\frac{\Pi^{\m\n}(p)}{\sqrt{p^2}}-\frac{W_J}{C_J^2+W_J^2}
\epsilon^{\m\n\rho}p_\rho\,.
\end{equation}
This does not depend on the gauge fixing parameter $\xi$ and
it is obtained from the initial two-point function (\ref{2ptJW}) by the
$S$-operation (\ref{JStransf}). 

\subsection{The $S$-operation on the two-point function of the 
energy momentum tensor}

In three dimensions there are two possible terms in the two-point function
of a symmetric traceless and conserved rank-2 tensor
\be\label{TTprop}
\langle T_{\mu\nu}(p)T_{\lambda\rho}(-p)\rangle=C_T
\frac{\Pi^{(2)}_{\mu\nu,\lambda\rho}(p)}{\sqrt{p^2}}
+W_T\Pi^{(1.5)}_{\mu\nu,\lambda\rho}(p)\,,
\end{equation}
or, in coordinate space
\be
\label{TTx}
\langle T_{\mu\nu}(x)T_{\lambda\rho}(0)\rangle=C_T
\Pi^{(2)}_{\mu\nu,\lambda\rho}(i\partial)\left(\frac{1}{2\pi^2x^{2}}\right) 
+W_T\Pi^{(1.5)}_{\mu\nu,\lambda\rho}(i\partial)\,\delta^{(3)}(x)\,.
\end{equation}
We have defined
\begin{eqnarray}
\Pi^{(2)}_{\mu\nu,\lambda\rho}(p)&=&\frac12\left[ \Pi_{\mu\lambda}(p) \Pi_{\nu\rho}(p)+
\Pi_{\mu\rho}(p) \Pi_{\nu\lambda}(p)-\Pi_{\mu\nu}(p) \Pi_{\lambda\rho}(p)\right]\,,\\
\Pi^{(1.5)}_{\mu\nu,\lambda\rho}(p)&=&  \frac14\left[\epsilon_{\mu\lambda\sigma}\Pi_{\nu\rho}(p)
+\epsilon_{\nu\lambda\sigma}\Pi_{\mu\rho}(p)
+\epsilon_{\mu\rho\sigma}\Pi_{\nu\lambda}(p)
+\epsilon_{\nu\rho\sigma}\Pi_{\mu\lambda}(p)\right]p^\sigma\,.
\end{eqnarray}
Note that the second term in (\ref{TTx}) is a conformally invariant
contact term special to
three dimensions. 
If we couple $T_{\mu\nu}$ to an external field $h_{\mu\nu}$ and
integrate out the elementary fields (assuming again an appropriate
Lagrangian formulation), we would like to
invert the induced $h_{\mu\nu}$ propagator. In order to do so, we must gauge
fix. A sufficiently general gauge fixing involves two arbitrary parameters $\xi_1$
and $\xi_2$ and is of the form (for clarity, we suppress
in the following 
indices that are not necessary) 
\begin{equation}
\label{TTgfix}
\langle TT\rangle\rightarrow
\langle TT\rangle_\xi=M=C_T\frac{\Pi^{(2)}}{\sqrt{p^2}}+W_T\Pi^{(1.5)}+p^3\left(
\xi_1 {\cal A}+\xi_2{\cal B}\right) \,,
\end{equation}
where it is convenient to define
\begin{eqnarray}
{\cal A}_{\mu\nu,\lambda\rho}&=&-\frac{3}{2p^4}\left( p_\mu
p_\nu-\frac13 p^2\eta_{\mu\nu}\right)\left( p_\lambda p_\rho-\frac13
p^2\eta_{\lambda\rho}\right)\,,\\ 
{\cal B}_{\mu\nu,\lambda\rho}&=&\frac{1}{2p^2}\left(p_\mu
p_\lambda\eta_{\nu\rho}+p_\mu p_\rho\eta_{\lambda\nu}+p_\nu
p_\lambda\eta_{\mu\rho}+p_\nu
p_\rho\eta_{\lambda\mu}\right)-\frac{4}{p^4}p_\mu p_\nu p_\lambda
p_\rho\nonumber\,. 
\end{eqnarray}
The inverse of (\ref{TTgfix}) may now be computed and we find
\be
p^6M^{-1}=\frac{C_T}{C_T^2+W_T^2}\frac{\Pi^{(2)}}{\sqrt{p^2}}
-\frac{W_T}{C_T^2+W_T^2}\Pi^{(1.5)}
+p^3\left( \frac{1}{\xi_1}{\cal A}+\frac{1}{\xi_2}{\cal B}\right)\,.
\end{equation}
This is the propagator of the field $h_{\m\n}$ which has zero scaling
dimension. If we define now a symmetric traceless and conserved
tensor in momentum space as
\begin{equation}
\hat{T}_{\mu\nu}=\Pi^{(1.5)}_{\mu\nu,\lambda\rho} h^{\lambda\rho}\,,
\end{equation}
we find that
\begin{equation}
\label{TTdual}
\langle
\hat{T}_{\mu\nu}\hat{T}_{\lambda\rho}\rangle=
\frac{C_T}{C_T^2+W_T^2}\frac{\Pi^{(2)}_{\mu\nu,\lambda\rho}(p)}
{\sqrt{p^2}}-\frac{W_T}{C_T^2+W_T^2}\Pi^{(1.5)}_{\mu\nu,\lambda\rho}(p)\,.
\end{equation}
We see that (\ref{TTdual}) does not depend on the gauge fixing
parameters and is of the form (\ref{TTprop}), but
with the coefficients transformed by $S\in SL(2,\bZ)$. 

\subsection{Generalization to higher-spin currents}

The computations presented in the previous  sections for $s=1,2$ can be
generalized to higher-spin  conserved currents. In three dimensions,
there are two possible conformally invariant tensor structures that may
appear in the momentum space two-point function of a spin-$s$ conserved
current (refer to the latter as $T_{(s)}$ and suppress indices)
\begin{equation}
\label{TT0}
\langle T_{(s)}T_{(s)}\rangle=C_s\frac{\Pi^{(s)}(p)}{\sqrt{p^2}}+W_s\Pi^{(s-1/2)}(p)\,.
\end{equation}
The first term in (\ref{TT0}) is the usual term (see for example
\cite{Tassos0}) that corresponds to an appropriately symmetrized and
traceless product of (\ref{Pimn}), while the second term is a
conformally invariant contact term. These terms have the schematic form
\begin{equation}
\label{spinstens}
\Pi^{(s)}\sim (\Pi^{(1)})^s\,,\,\,\,\,\, \Pi^{(s-1/2)}\sim (\Pi^{(1)})^{s-1}\Pi^{(0.5)}\,,
\end{equation}
where by $\Pi^{(0.5)}, \Pi^{(1)}$ we mean
$\Pi^{(0.5)}_{\mu\nu}=\epsilon_{\mu\nu\lambda}p^\lambda$ and
$\Pi^{(1)}_{\mu\nu}=\Pi_{\mu\nu}$. The index structure of each tensor in
(\ref{spinstens}) is {\em uniquely} determined by symmetry  and tracelessness,
while the requirement of canonical dimensions $\Delta=s+1$ also implies
conservation. By coupling to an external field $h_{(3-s)}$ and gauge
fixing in order to invert its propagator, one finally arrives at the
gauge independent two-point function
\begin{equation}
\langle \hat{T}_{(s)}\hat{T}_{(s)}\rangle=
\frac{C_s}{C_s^2+W_s^2}\frac{\Pi^{(s)}(p)}{\sqrt{p^2}}-
\frac{W_s}{C_s^2+W_s^2}\Pi^{(s-1/2)}(p)\,,
\end{equation}
for the dual current
\be 
\hat{T}_{(s)}= \Pi^{(s-1/2)} h_{(3-s)}\,.
\ee

\section{Double-trace deformations  and the $S$-operation}

As a first step towards understanding our results so far, we ask if the
$SL(2,\bZ)$ action, and in particular the $S$-operation, have an
underlying dynamics and whether this dynamics is generic in
three-dimensional CFTs. Rather surprisingly, we find that dynamically
induced duality transformations are common in three-dimensional CFTs.
Moreover, the underlying dynamics appears to be rather generic and
amounts to double-trace deformations. Indeed, we are able to explicitly
demonstrate that the $S$-operation on spin-1 and spin-2 conserved
currents can be induced by certain irrelevant double-trace deformations,
under the assumption that these lead to well-defined UV fixed-points.
Therefore, we see the formation of an interesting pattern on the space
of three-dimensional CFTs.

\subsection{A precursor to the $S$-operation}

To set the stage, we recall here some of the salient properties of two
well-known three-dimensional theories with non-trivial (large-$N$) fixed
points: the critical $O(N)$ Vector model and Gross-Neveu models. The
corresponding (Euclidean) actions are
\bea
\label{ONGN}
{\cal L}_{O(N)} &=&\int d^3x \left[\frac{1}{2}\left(\partial \phi\right)^2 +\frac{1}{2}
\sigma\left(\phi^2-\frac{1}{g}\right)\right] \,,\\
{\cal L}_{GN} &=&-\int
d^3x\left[\bar{\psi}\slash\!\!\!\partial\psi+\frac{G}{2}\left(\bar{\psi}\psi\right)^2
\right]\,,
\eea
for the $N$-component scalar $\phi(x)$ and Majorana fermion $\psi(x)$.
The auxiliary scalar $\sigma(x)$ enforces the constraint in the $O(N)$
Vector model. Both models are renormalizable in the $1/N$ expansion and
have non-trivial large-$N$ fixed points corresponding to strongly
coupled CFTs where the elementary fields $\phi(x)$ and $\psi(x)$ 
acquire anomalous dimensions of order $1/N$. Most importantly, and this
is the key to the $1/N$ renormalizability of the models, the composite
operators $\phi^2(x)$ and $(\bar{\psi}\psi)(x)$ receive large
corrections to their scaling dimensions.

We may interpret this latter phenomenon as a duality transformation in
the following sense. In a three-dimensional CFT all quasi-primary
operators must transform under unitary irreducible representations (UIR)
of $SO(3,2)$. For example, the bilinear leading-twist\footnote{Operators
containing more than two $\phi^a(x)$'s are {\it higher-twist}.} 
operators in the $O(N)$-singlet sector of the free field theory limit of
the $O(N)$ Vector model, realize the UIRs denoted by $\chi
=[\Delta_0,s]$ where $\Delta_0$ is the {\it canonical} scaling dimension
and $s$ the total spin. In particular, we note that the composite
operator $\phi^2(x)$ realizes the UIR $[1,0]$ \cite{Fronsdal}. A similar
classification must also hold true for the quasi-primary operators of
the corresponding sector at the large-$N$ fixed point of the model.
Explicit calculations then show that almost all leading-twist
quasi-primary operators of the non-trivial CFT realize UIRs of the form
$[\Delta_0+o(1/N),s]$ \cite{Ruhl,Tassos2}.  The important exception is
connected with the large-$N$ realization of the operator $\phi^2(x)$. In
the non-trivial CFT, the UIR $[1,0]$ does not exist and instead one
finds the  UIR $[2+o(1/N),0]$. Therefore, in the strict large-$N$ limit
this particular sector of the operator spectrum in the non-trivial CFT
is obtained from the spectrum of the free CFT by changing the UIR
$[1,0]$ to the UIR $[2,0]$. As is well known, these two UIRs are
equivalent and are exchanged by a Weyl reflection \cite{Koller,Dobrev}
in $SO(3,2)$.\footnote{In general, a Weyl reflection exchanges the UIRs
$\chi =[\Delta,s]$ and $\tilde{\chi}=[d-\Delta,s]$. These are called
{\it shadow} UIRs and have the same Casimirs. The conformally invariant
two-point function corresponding to the UIR $\chi$ is then defined as an
intertwiner between the UIRs $\chi$ and $\tilde{\chi}$. The symmetry
between the two is broken in a given CFT by the Ward identities which
pick one UIR to be ``elementary".} In that sense, the non-trivial
large-$N$ fixed point of the $O(N)$ Vector model may be viewed as the
result of a `duality' transformation on the spectrum of the free CFT. In
exactly the same way, a particular sector in the spectrum of the
non-trivial fixed point of the $O(N)$ Gross-Neveu model is obtained, in
the strict large-$N$ limit, by implementing a duality operation on the
spectrum of the free fermionic CFT. In that case one changes the UIR
$[2,0]$ that is realized by the composite operator $(\bar{\psi}\psi)(x)$
to its shadow UIR $[1,0]$.

The above duality operation may be viewed as a precursor to the
$S$-operation acting on two-point functions of scalars. Moreover, we
have a good understanding of the dynamics underlying this operation
which is similar in both models. Indeed, we know that the non-trivial
fixed-point in each model is reached under the influence of a certain
double-trace deformation. In the case of the $O(N)$ Vector model, the
deformation is relevant and the corresponding fixed-point is in the IR,
while in the case of the Gross-Neveu model the deformation is irrelevant
and reveals the existence of a non-trivial UV fixed point. Let us see
how such generic double-trace deformations induce the duality operation
that we have mentioned. Consider a three-dimensional Euclidean CFT with
elementary fields $\phi$ and partition function
\be
\label{PF}
{\cal Z} = \int ({\cal D}\phi)e^{-S(\phi)}\,.
\ee
Consider now a scalar composite operator ${\cal O}(x)$ with canonical
dimension $1/2<\Delta<3/2$. 
We deform the action by 
\be
\label{deform}
-\frac{f}{2}\int d^3x\,\,{\cal O}^2(x)\,,
\ee
and ask for the two-point function of ${\cal O}(x)$ at
the IR fixed point where the deformation (\ref{deform}) presumably
leads the theory (\ref{PF}). We proceed
by a direct calculation 
\bea
\label{OOf}
\langle {\cal O}(x_1){\cal O}(x_2)e^{\frac{f}{2}\int d^3x\,{\cal
O}^2(x)}\rangle &=&\langle {\cal O}(x_1){\cal O}(x_2)\rangle_f
\nonumber \\
&=& \langle{\cal O}(x_1){\cal O}(x_2)\rangle_0 +\frac{f}{2}\int d^3x
\langle {\cal O}(x_1){\cal O}(x_2){\cal O}^2(x)\rangle_0 \nonumber \\
&+&\frac{f^2}{8}\int d^3x d^3y\langle {\cal O}(x_1){\cal O}(x_2){\cal
O}^2(x){\cal O}^2(y)\rangle_0 +\ldots\,.
\eea
We can use the OPE of ${\cal O}(x)$ with itself to calculate the
correlation functions in (\ref{OOf}). Moreover, we assume now the
existence of a suitable large-$N$  expansion such that the leading $N$
contribution comes from the two-point function $\langle {\cal O}{\cal
O}\rangle$. Then, taking into account the combinatorics we find
\be
\label{OOf1}
\langle{\cal O}(x_1){\cal O}(x_2)\rangle_f =\langle{\cal O}(x_1){\cal O}(x_2)\rangle_0
+ f\int d^3x\langle{\cal O}(x_1){\cal O}(x)\rangle_0\langle {\cal
O}(x){\cal O}(x_2)\rangle_f+\cdots\,,
\ee
where the dots contain terms subleading in $1/N$ that we drop. Denoting
the momentum space two-point function by $Q(p)$, we obtain from
(\ref{OOf1})
\be
\label{Qfmomentum}
Q_f(p)=\frac{Q_0(p)}{1-fQ_0(p)}\,.
\ee
The assumption of the existence of a non-trivial fixed-point 
enters now in a crucial way. Indeed, the coupling has dimension
$[mass]^{3-2\Delta}$ and sets the scale of physical processes. To
study the IR behavior of (\ref{Qfmomentum}) we need to assume that $f$
can be made finite by renormalization and expand for small momenta $p$
as
\be
\label{Qf}
f^2Q_f(p)=-\frac{f}{\left(1-\frac{1}{fQ_0(p)}\right)}
=-f-Q_0^{-1}(p) +\cdots\,,\,\,\,\,\,
Q_0(p)\approx \frac{1}{p}\gg 1\,. 
\ee
This is the properly normalized two-point function of the operator
${\cal O}(x)$ for mass scales less than the scale set by the
renormalized coupling. The minus sign in the second term on the rhs of
(\ref{Qf}) guarantees the positivity of the $x$-space two-point function
in the IR, while the first term gives a conformally non-invariant
contact term in $x$-space and should be dropped. We see then that the
effect of the deformation (\ref{deform}) is the duality operation
$Q_0(p)\rightarrow f^2\,Q_f(p)\approx -Q_0^{-1}(p)$, or in an algebraic
sense to change the UIR $[1,0]$ to $[2,0]$. This is very similar to the
``first-half'' of the  $S$-operation (e.g. going from $J^\m$ to $A_\m$
for $W=0$), discussed in the previous section.

In a similar fashion we may consider theories having a scalar operator
$\Psi(x)$ with dimension $3/2< \Delta <3$ and deform by the irrelevant
double-trace deformation
\be
\label{irrdeform}
-\frac{G}{2}\int d^3x\Psi(x)\Psi(x)\,.
\ee
In this case one expects that the initial theory may be
the IR limit of a non-trivial UV fixed point, as in the explicit
example of the Gross-Neveu model. As before, we assume a large-$N$
expansion for the correlation functions involving $\Psi(x)$ and
$\Psi^2(x)$ and by a direct calculation we obtain
\be
\label{Psimom}
{\cal J}_G(p) =\frac{{\cal J}_0(p)}{1-G{\cal J}_0(p)}\,,
\ee
where ${\cal J}(p)$ denotes the momentum space two-point function of
$\Psi(x)$. The coupling $G$ has negative mass dimension and as before
we have to assume that it can be made finite by renormalization. Then
we can study the large momentum behavior of the properly normalized
two-point function of $\Psi(x)$ for momenta much larger than the scale
set by the renormalized coupling as
\be
\label{JG}
G^2{\cal J}_G(p)=-\frac{G}{\left(1-\frac{1}{G{\cal J}_0(p)}\right)}
=-G-{\cal J}_0^{-1}(p) +\cdots\,,\,\,\,\,\,
-{\cal J}_0(p)\approx \sqrt{p^2}\gg 1\,.
\ee
As before, up to a contact term this gives a positive definite
$x$-space UV two-point function. Therefore the irrelevant deformation
(\ref{irrdeform}) produces the duality operation ${\cal
J}_0(p)\rightarrow G^2{\cal J}_G(p) \approx -{\cal J}_0^{-1}(p)$ and is
also very similar to the ``first-half'' of the 
$S$-operation of the previous section.

An important property of the
UIRs $[1,0]$ and $[2,0]$ is the fact that they are both
above the unitarity bound\footnote{The unitarity bound for the scaling
dimensions in $SO(3,2)$ is $\Delta\geq 1/2$ for spinless irreps and
$\Delta\geq s+1$ for irreps with spin $s\geq 1$.} for UIRs of
$SO(3,2)$ \cite{Evans}. This is exceptional, as the same is no longer
true if one wished to perform the same duality operation on most of
the UIRs of the free CFT. In particular, all the shadows of the UIRs
$[s+1,s]$ that correspond to the infinite set of conserved currents in
the free CFT fall below the unitarity bound.  Nevertheless, there
exist explicit examples of three-dimensional theories where 
the puzzle is apparently resolved. These examples correspond to
fermionic theories with action of the generic form 
\be
\label{Thir} 
S =-\int d^3x
\left[\bar{\psi}\slash\!\!\!
\partial\psi+\frac{F}{2}(\bar{\psi}\gamma_\m\psi)(\bar{\psi}\gamma_\m\psi)
\right]\,,
\ee
for the $N$-component Majorana spinor $\psi(x)$. The conserved $U(1)$
current is $J_\m(x)=\bar{\psi}(x)\gamma_\m\psi(x)$ and while the
interaction in (\ref{Thir}) appears non-renormalizable (irrelevant) by
power counting, such models are renormalizable in the $1/N$ expansion
and also have non-trivial UV fixed points that correspond to gauge CFTs
\cite{Parisi,Hands,Anselmi,Kapustin,Witten}. In contrast to the
Gross-Neveu model the bare coupling constant $F$ is not renormalized.

The puzzle with the duality operation is resolved in an interesting way
in these models \cite{FP}. The shadow operator of the $U(1)$ current must be
regarded as a gauge field. Indeed, this is the main finding of studies
in these models and has led to the understanding that the UV fixed-point
is related to conformal QED$_3$. Gauge fields are not required to
realize UIRs of the conformal group;  only gauge invariant quantities
do. On the other hand, given a three-dimensional gauge field $A_\m(x)$,
one can construct the gauge invariant conserved current
\be
\label{giJ}
\hat{J}^\m(x) =\epsilon^{\m\n\rho}\partial_\n A_\rho(x)\,.
\ee
which realizes the standard UIR $[2,1]$ of $SO(3,2)$. The
lengthy discussion above begins to clarify. The duality operation on
the spin-1 current of the free theory corresponds to the ``first-half''
of the $S$-operation and will lead to the gauge field. The construction
(\ref{giJ}) provides the ``second-half'' of the $S$-operation and one
finally gets the spin-1 current of the non-trivial theory. The
underlying dynamics is a double-trace deformation. We demonstrate this in
the next subsection.

\subsection{Double-trace deformations and the $S$-operation on the
two-point function of the spin-1 current}

Consider a three-dimensional CFT with a $U(1)$ current $J_\m(x)$ having
momentum space two-point function (\ref{2ptJW}). We  perturb the theory
by the irrelevant deformation
\be
\label{deformJ}
-\frac{f}{2}\int\rmd^3x J_{\m}(x)J_\m(x)\,.
\ee
To proceed with a calculation similar to (\ref{OOf}) we assume a
$1/N$ expansion of the correlation functions involving $J_\m(x)$ and
$J_\m^2(x)$, with leading terms coming from the two-point function of
$J_\m(x)$. We then obtain 
\be
\label{JJf}
\left[{\delta^\m}_{\rho}-f{J_0^{\m}}_{\rho}(p)\right]
J_f^{\rho\n}(p)=J_0^{\m\n}(p)\,.
\ee
As before, the existence of a non-trivial UV fixed point implies that
the solution for large momenta is of the form
\be
f^2\,J_f^{\rho\n}(p)\simeq -f\,{\delta^{\rho\n}}-{(J_0^{-1})^{\rho\n}}+\ldots\,,
\ee
which is similar to (\ref{Qf}) and (\ref{JG}). In the present case,
however, we need to gauge fix so that $J_0^{-1}$ exists. Alternately, we
may simply write the ansatz
\be\label{Jinverse1}
f^2\,J_f^{\mu\nu}(p)=\hat C_J\frac{\Pi^{\mu\nu}(p)}{\sqrt{p^2}}+\hat
W_J\epsilon^{\mu\nu\rho}p_\rho 
\ee
and using (\ref{JJf}) we easily find\footnote{Similar formulas were derived also in \cite{BD}.}
\begin{eqnarray}
\hat C_J&=&\frac{f^2[C_J+fp(C_J^2+W_J^2)]}{1+2fpC_J+f^2p^2(C_J^2+W_J^2)}\,,\\
\hat W_J&=&\frac{f^2\,W_J}{1+2fpC_J+f^2p^2(C_J^2+W_J^2)}\,.
\end{eqnarray}
We want to study this
for large momenta (in the UV), assuming at the same time a finite
critical coupling $f$. One obtains,
\begin{eqnarray}
\label{hCJ}
\hat C_J&\sim& \frac{f}{\sqrt{p^2}}-\frac{1}{p^2}\frac{C_J}{C_J^2+W_J^2}+\ldots\\
\label{hatWJ}
\hat W_J&\sim& \frac{1}{p^2}\frac{W_J}{C_J^2+W_J^2}+\ldots\label{hat WJ}
\end{eqnarray}
The first term on the rhs of (\ref{hCJ}) gives a
non-conformally-invariant contact term which is dropped. The two-point
function (\ref{Jinverse1}) can be viewed as the properly normalized
two-point function of a vector operator with dimension 1 and to avoid
problems with the unitarity bound we must assume that this is a gauge
field.  Then the two-point function of the associated gauge invariant
conserved current, which we write in momentum space as  $\hat{J}^{\m}(p)
={i}\epsilon^{\m\n\rho}p_\n A_\rho(p)$, is found to be
\be\label{Jhat}
\hat
J^{\mu\nu}(p)=\frac{C_J}{C_J^2+
W_J^2}\frac{\Pi^{\mu\nu}(p)}{\sqrt{p^2}}- 
\frac{W_J}{C_J^2+W_J^2}\epsilon^{\mu\nu\rho}p_\rho\,.  
\ee
We see that (\ref{Jhat}) is obtained from (\ref{2ptJW}) by the
$S$-operation (\ref{JStransf}).

\subsection{Double-trace deformations and the $S$-operation on the
two-point function of the energy momentum tensor}

That our previous calculations work out is perhaps not surprising, since
our lengthy discussion of the various well-known three-dimensional
CFTs unveils a similar underlying dynamics. What is rather surprising
is that a similar double-trace deformation appears also to be the
underlying dynamics of the $S$-operation on the two-point function of
the energy momentum tensor. Consider the irrelevant deformation
\be
\label{deformTT}
-\frac{g}{2}\int d^3x\ T_{\m\n}(x)T_{\m\n}(x)\,.
\ee
Under the same assumptions as before, namely the existence of a
suitable large-$N$ limit for correlation functions and a non-trivial UV
fixed point, following similar calculations we obtain
\be
\label{basicT}
\left[\delta_{\m\a}\delta_{\n\b}-gT^0_{\m\n ,\a\b}(p)\right]T^g_{\a\b
,\rho\sigma}(p) =T^0_{\m\n ,\rho\sigma}(p)\,.
\ee
This gives for large momenta
\be
\label{Tgexp}
g^2\,T^g_{\m\n
,\rho\sigma}(p) \approx g^2\,\delta_{\m\n}\delta_{\rho\sigma}
-(T^{-1}_0)_{\m\n,\rho\sigma}(p)+\cdots\,.
\ee
Again we need to gauge fix such that $T_0^{-1}$ exists, but with our
previous experience we try an ansatz of the form
\be
\label{Tgansatz}
g^2\,T^g_{\m\n ,\rho\sigma}(p)=\hat{C}_T
\frac{\Pi^{(2)}_{\mu\nu,\lambda\rho}(p)}{\sqrt{p^2}} 
+\hat{W}_T\Pi^{(1.5)}_{\mu\nu,\lambda\rho}(p)\,,
\ee
After some lengthy but straightforward algebra we obtain
\bea
\label{hatCT}
\hat{C}_T &=&\frac{g^2[C_T-gp^3(C^2_T+W_T^2)]}{1+(gp^3)^2(C_T^2 +W_T^2)
-2gp^3C_T}\,,\\
\label{hatWT}
\hat{W}_T &=& \frac{g^2\,W_T}{1+(gp^3)^2(C_T^2 +W_T^2)
-2gp^3C_T}\,,
\eea 
in complete analogy with (\ref{hCJ}) and (\ref{hatWJ}). The
result is a two-point function of a dimension zero symmetric and
traceless tensor $h_{\m\n}$ which, as before, we must require to be a gauge field. Gauge symmetry in this case may be local
diffeomorphisms and $h_{\m\n}$ may be viewed as the symmetric
traceless part of the metric tensor in three-dimensional gravity. 
Then, the gauge invariant symmetric, traceless and conserved tensor
of dimension three
\be
\label{Tdual}
\hat{T}_{\m\n}(p)=i\Pi^{(1.5)}_{\m\n ,kl}h_{kl}(p)\,,
\ee
has two-point function exactly of the form (\ref{TTdual}). Clearly, it would be of interest to find an explicit theory where this is implemented.

One can also
straightforwardly show that an irrelevant 
double-trace deformation of a similar form induces the $S$-operation on
the two-point functions of higher-spin conserved currents. We leave this for future work.

\section{The Bulk View}

Witten has suggested \cite{Witten} that the action of $SL(2,\bZ)$ on the
correlators of abelian currents can be understood as the holographic
image of electromagnetic duality of a $U(1)$ gauge theory on AdS$_4$.
Specifically, he suggested that the S-operation corresponds to a
choice of boundary condition. Given our results for the energy momentum
tensor two-point function, we can ask if at least in some approximation,
there is a natural place for $SL(2,\bZ)$ in a four-dimensional gravity
theory. Let us first consider the case of gravity in the $AdS_4$ bulk.
It is convenient to write the Einstein-Hilbert action in
MacDowell-Mansouri form \cite{MM}. To do so, we formally introduce an
$SO(3,2)$ connection ${\omega^A}_B=-{\omega^B}_A$ with curvature ${{\cal
R}^A}_B=d{\omega^A}_B+{\omega^A}_C\wedge {\omega^C}_B$. An appropriate
action is of the form  \cite{SW}
\begin{equation}
\label{actionMM}
I_{MM}=-\frac{1}{4L}\,\int_{\cal M} V^A {\cal R}^{BC}{\cal
R}^{DE}\epsilon_{ABCDE}\,,\,\,\,\,\epsilon_{-10123}=1\,, \,\,\,\,\eta_{AB}=(-+++-)\,,
\end{equation}
where $V^A$ is a $SO(3,2)$ vector satisfying $V^A V_A=-L^2$. This is
actually the defining condition for AdS$_4$ with radius $L$ embedded into a
five-dimensional space with metric $\eta_{AB}$. The gauge choice
$V^{-1}=L, V^a=0$, $a=0,1,2,3$ splits the connection into the
$SO(3,1)$ connection and the vierbein as
\begin{equation}
\label{veirbein}
{\omega^A}_B\rightarrow \,\,\,\,\,{\omega^a}_b\,\,,\,\,
{\omega^a}_{-1}=\frac{1}{L}e^a\,. 
\end{equation}
The curvature then splits into two pieces
\begin{eqnarray}
{{\cal R}^a}_b&=&{R^a}_b+ \frac{1}{L^2}e^a\wedge e_b\,,\\
{{\cal R}^a}_{-1}&=&\frac{1}{L}de^a+\frac{1}{L}{\omega^a}_b\wedge e^b= T^a\,,
\end{eqnarray}
the latter being the torsion.
The MacDowell-Mansouri action then reduces to
\begin{eqnarray}
\label{actionMM1}
I_{MM}=
&&-\frac{1}{2L^2}\int_{\cal M} e^a\wedge e^b\wedge
R^{cd}\epsilon_{abcd} -\frac{1}{4L^4}\int_{\cal M} 
e^a\wedge e^b\wedge e^c\wedge e^d\epsilon_{abcd} +
\frac{1}{2}\int_{\cal M} tr\
R\wedge\tilde R\nonumber \\ 
&&\hspace{2cm} =\frac{1}{L^2}\int_{\cal M} d^4x\sqrt{-g}\
\left(R+\frac{6}{L^2}\right)-8\pi^2\chi({\cal M})\,, 
\end{eqnarray}
where $\tilde R^{ab}=\frac12\epsilon^{abcd}R_{cd}$ and $\chi({\cal M})$
is the Euler character. (Of course we are being notationally loose here:
AdS is not compact, but we will discuss boundary terms presently). In
fact, the Euler character is of little interest for the present
discussion: it is topological, and the quadratic boundary terms that it
creates are divergent and are removed by boundary counterterms. On the
other hand, it is well-known that the Einstein-Hilbert term leads to the
standard $C_T$ term in the two-point function of the energy momentum
tensor in the boundary theory \cite{LT}.

There is however another topological term that is of interest to our
discussion. It is the (bulk part of the) Pontryagin class
\begin{equation}
P(M)=-\frac{1}{8\pi^2}\int_{\cal M} {{\cal R}^A}_B {{\cal
R}^B}_A=-\frac{1}{8\pi^2}\int_{\cal M} Tr\ R\wedge R\,, 
\end{equation}
the latter equality holding up to torsion terms on the boundary. As we
stated, this term is topological and gives a boundary contribution
\begin{equation}
\label{PM}
P({\cal M})=-\frac{1}{8\pi^2}\int_{\partial {\cal M}} Tr \left(\omega\wedge
d\omega+\frac23\omega\wedge\omega\wedge\omega\right)\,.
\end{equation}
To evaluate (\ref{PM}) we expand the metric to linearized level around
the AdS background 
\begin{equation}
ds^2=\bar g_{ab}dy^a dy^b=\frac{L^2}{r^2}\left(
dr^2+\eta_{\mu\nu}dx^\mu dx^\nu\right)\,,\,\,\, \m ,\n =1,2,3
\end{equation}
as $g_{ab}=\bar{g}_{ab}+r^{-2}\,h_{ab}$. Substituting
this back to (\ref{PM}) we obtain
\begin{equation}
P({\cal M})=-\frac{1}{16\pi^2}\int d^3x\ \epsilon^{\mu\nu\lambda} 
{{h_\mu}^\epsilon}_{,\sigma}  \left(
{{h_\lambda}^\sigma}_{,\epsilon}-{h_{\lambda\epsilon}}^{,\sigma}\right)_{,\nu}\,,
\end{equation}
where $h^{\m\n}(\bar{x})$ is the boundary value of the metric
fluctuation and indices have been raised and lowered with
$\eta_{\m\n}$. In momentum space, we find 
\begin{equation}
P({\cal M})=\frac{i}{16\pi^2}\int \frac{d^3k}{(2\pi)^3}
h^{\mu\epsilon}(k)\Pi^{(1.5)}_{\mu\epsilon,\lambda\sigma} (k)
h^{\lambda\sigma}(-k) \,.
\end{equation}
Clearly, this gives the contribution to the energy momentum tensor two-point
function proportional to $W_T$.  

The above results lead us to consider the following Euclidean bulk action 
\begin{equation}
\label{selfdual}
S_{bulk}=\frac{1}{4\pi}\left[ \frac{4\pi}{g_N^2}\int_{\cal M} Tr\ {\cal
R}\wedge \tilde {\cal R}+i\frac{\theta_N}{2\pi}\int_{\cal M} Tr\ {\cal
R}\wedge {\cal R}\right]+\int_{\partial {\cal M}} {\cal L}_{c.t.}\,, 
\end{equation}
where we have introduced a dimensionless coupling $g_N$ which in the
standard normalization of the Einstein-Hilbert action is
\begin{equation}
\frac{4\pi}{g^2_N}=\frac{L^2}{8G_4}\,,
\end{equation}
and a theta-angle $\theta_N$.
The action (\ref{selfdual}) expanded around the AdS$_4$ background gives
rise to the boundary two-point function (\ref{TTprop}). The parameters
$C_T$ and $W_T$ depend respectively on $g_N$ and $\theta_N$. 

Now we see that the shift of $\theta_N\to\theta_N+2\pi$ induces a
shift of $W_T$ in the boundary theory: this is the T-operation.  
On the other hand, the action has been deliberately written in a form which is
reminiscent of gauge theory. In particular, we can also write the action in the form 
\begin{equation}
\label{selfdual1}
\frac{1}{4\pi}\sum_\pm \tau_\pm Tr\ {\cal R}^\pm\wedge {\cal R}^\pm\,,
\end{equation}
if we introduce 
\begin{eqnarray}
{\cal R}^\pm&=&\frac12\left( {\cal R}\pm\tilde {\cal R}\right)\,,\\
\tau_\pm &=& \frac{4\pi}{g_N^2}\pm i\,\frac{\theta_N}{2\pi}\,.
\end{eqnarray}
We see now that a least at the linearized level (i.e. if we consider
only two-point functions in the boundary) one expects that the
$S$-transformation in the boundary is induced by an appropriate
generalization of electric-magnetic duality transformation for the
bulk gravity. 

The generalization of our results for higher spins now looks
straightforward at the linearized level. One may start from the action
written down by Vasiliev in \cite{Vasiliev2} and add to it the
appropriate generalization of the Pontryagin term. It appears then
possible that one can arrive at an action that can be written in the
form of (\ref{selfdual1}), but we leave this discussion for the
future.

\section{Discussion}

There is a number of potentially interesting implications of our
results. On a broader level, they indicate an intimate
relation between $SL(2,\bZ)$ duality and the holography of free
field theories. In this context, it would be of interest to find
examples where the sort of duality 
that we have discussed here for spin $\geq 2$ is implemented. Certain
Higher-Spin theories that have been constructed on AdS$_4$ are a
natural place for such studies.

One can
also speculate as to whether or not there are theories for which the
duality operates beyond the linearized level. 
For this one should study three-point functions of spin-1,
spin-2 and perhaps higher-spin currents in order to understand whether
there exist some natural discrete action in three-dimensional CFTs
beyond the linearized level. Since conformal invariance fixes the form
of three-point functions up to a number of constant parameters, it is
possible to follow the calculations in this work and evaluate the
three-point function of the dual (non-abelian) spin-1, spin-2 and
higher-spin currents.

An intriguing and potentially far reaching implication of our results is
the possibility that many known three-dimensional CFTs can provide
information for quantum field theory on (A)dS$_4$, even without the
involvement of supersymmetry. At this point we note that the parameter
$N$ of the boundary CFTs is related to the Planck length of the bulk
theory. In this sense it is perhaps not inconceivable that the wealth of
statistical physics  phenomena could be directly related to as yet
undiscovered phenomena of quantum fields (or strings, membranes), on
(A)dS$_4$. As an example, one might wonder whether black hole-like
configurations in the bulk are related to thermal effects in the
boundary.

\acknowledgments 

We thank L. Andrianopoli, L. Alvarez-Gaum\'e, E. Floratos, A. Kovner, C. A. L\"utken,
D. Minic, Y. Oz, M. Porrati, P. Vanhove and M. Vasiliev for useful
discussions. The work of RGL is supported in part by 
the U.S. Department of Energy under contract DE-FG02-91ER40677. RGL
would also like to thank the Aspen Center for Physics for a
stimulating environment where some of this work was carried out.


\end{document}